\documentclass[prl,twocolumn,floatfix,showpacs]{revtex4}
\usepackage{amsmath}
\usepackage{epsfig}
\usepackage{bm}

\begin{document}

\title{A Physical Theory of the Competition that Allows HIV to Escape from
the Immune System}
\author{Guanyu Wang$^{1,2}$ and Michael W. Deem$^1$}
\affiliation{\hbox{}$^1$Department of Bioengineering and Department of Physics \&
Astronomy, Rice University, Houston, TX 77005--1892, USA \\
\hbox{}$^2$Department of Physics, George Washington University, Washington,
D.C. 20052, USA}

\begin{abstract}
Competition within the immune system
may degrade immune control of viral infections. We formalize
the evolution that occurs in both HIV-1 and the immune system quasispecies.
Inclusion of
competition in the immune system leads to a novel balance between the
immune response and HIV-1, in which the eventual outcome is HIV-1 escape
rather than control. 
The analytical model reproduces the three stages of HIV-1 infection.
We propose a vaccine
regimen that may be able to reduce competition between T cells, potentially
eliminating the third stage of HIV-1.
\end{abstract}

\maketitle


Our immune system is highly effective in suppressing most viral infections,
due to the many different T cells that exist in the repertoire of one
person. While many different T cells can
recognize a virus, only
those of highest affinity respond in large numbers and participate in
eliminating the virus. One limitation of our immune system stems from the
competition among T cells of similar specificity for the virus. For a series
of discrete infections over time, competition in the immune system is
associated with the phenomenon of deceptive imprinting \cite{Kohler}, or
original antigenic sin, and has been characterized by a random energy model
\cite{Deem}. Original antigenic sin is the tendency for memory immune cells
produced in response to a first viral infection to suppress the creation of
new immune cells in response to a second infection with a related strain.
Moreover, while they are used, these memory immune cells may not be optimal
for control of this second, different viral strain. Another form of
competition in the immune system occurs when several viral strains
simultaneously infect one person. In this case, the T cells compete to
recognize the different strains, and recognition of all strains may not be
uniformly effective. This immunodominance of one strain over others means
that the immune response to multiple infections is not a simple
superposition of the responses to each individual infection
\cite{dengue}.

\begin{figure}[t!]
\begin{center}
\epsfig{file=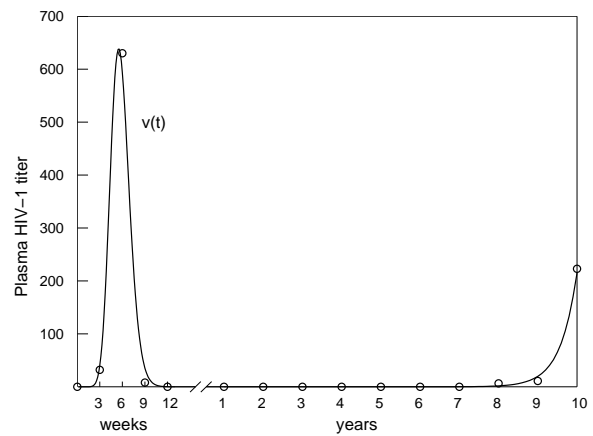,width=0.9\columnwidth,clip=}
\end{center}
\caption{The entire time course of HIV-1 infection. The circles are clinical
data from \protect\cite{Pantaleo}. The solid line is generated by our model.}
\label{fig:timecourse}
\end{figure}

For a chronic infection with a mutating virus, the effects of competition
are more subtle. Since the infection changes slowly over months or years,
there may be no apparent original antigenic sin, as the immune system slowly
updates itself in an attempt to clear the virus. If the virus evolves mainly
through point mutation, the mutants will be only slightly different from
their parents and can still be eliminated by the memory T cells. Moreover,
whenever a sufficiently new viral strain emerges through the accumulation of
point mutations that cannot immediately be cleared by the immune system,
there is enough time for the immune system to create new T cells against the
new strain since the infection is chronic. Therefore, competition might
represent a relatively small weakness of the immune system against chronic
infections. However, when facing HIV-1, any trivial weakness may become the
immune system's Achilles' heel. A typical HIV-1 infection is characterized
by three phases. See Fig. \ref{fig:timecourse}. 
The viral load initially
increases dramatically in the first viremia, followed by a sharp decline due
to immunological control. Thereafter, a long period of clinical latency
lasts for 6--10 years with small viral load. Finally, the HIV-1 viral load
climbs up again in the final viremia, in the onset of AIDS. Most
modeling works focus on the first viremia (e.g., \cite{Perelson,
Rouzine}).  In \cite{Kamp}, the second and third stages were
modeled but without the first stage. The entire time course has been
difficult to reproduce faithfully by a unified mathematical model, probably
due to the disparate dynamics of the three phases. To describe each phase, a
different set of parameters has been given. The three stage dynamics
is reproduced by computational models in \cite{Hershberg, Rita}, but
the mapping between parameters and physiology remains to be fully established
in those models \cite{strain}.

In this letter, we develop an analytical, physical theory
of the immune response that
explains HIV-1 pathogenesis. A key concept in the model is a population of
viruses with different sequences that arise due to viral evolution, termed
the viral quasispecies. Since HIV-1 and the immune system are well-matched
competitors, the viral quasispecies and its dual, the lymphocyte
quasispecies, should be equally well matched. These considerations lead to a
theory with a large number of dynamical equations, which is reduced to a low
dimensional model by symmetry. The model fits well with the standard HIV-1
data of the entire infection time course, and the parameters are all
biologically reasonable. A novel equilibrium between the immune system and
virus quasispecies is found, which, instead of fixing the viral load, favors
viral escape.

We let $V_{i}$ ($i=0,1,\cdots I-1$) denote the HIV-1 quasispecies $i$. We
let $v_{i}$ denote the concentration, or load, of $V_{i}$. We let $%
v=\sum_{i=0}^{I-1}v_{i}$ denote the total viral load. We let $X_{i}$ denote
the T cell, or effector T lymphocyte, that is specific for $V_{i}$. We let $%
x_{i}$ denote the concentration of this T cell, and we let $%
x=\sum_{i=0}^{I-1}x_{i}$ denote the total T cell concentration. We let $%
k_{ji}$ denote the affinity between the T cell $X_{j}$ and the virus $V_{i}$%
. As a fairly generic assumption, we take $k_{ij}=\exp ( -\beta
d_{ij}) $, where $d_{ij}=d_{ji}$ is the Hamming distance between $V_{i}
$ and $V_{j}$. Since roughly two amino acid substitutions still permit
recognition, we expect $\beta $ to be on the order of $1/2$.
The $k_{ij}$ ($i\neq j$)
represent cross-reactivities, which are generally beneficial for the immune
system because they offer additional killing. However, the side effect of
cross-reactivity is that it induces competition when the T cells
expand in concentration during the activation phase of the immune response.

We first hypothesize an ideal immune system that has the following
properties: (i) There is no competition among the different T cells. This
allows for the free development of effector T cells whose total
concentration will reach its maximum capacity, $x = \sum_{i=0}^{I-1} x_i = 1$%
. For a realistic system, we will find $x<1$ due to competition. (ii) There
is no bias toward any viral quasispecies in generating its corresponding
effector cells. Therefore one has $x_{i}/x=v_{i}/v$. Since $x=1$, one has $%
x_{i}=v_{i}/v$.

The real immune system and the competition induced in the response to HIV-1
are modeled as follows. Compared to cytotoxic T cell lymphocytes (CTLs),
antibodies are much less effective in controlling HIV-1. Therefore, we focus
on the CTL responses, and a virus is accordingly identified by its epitope
that stimulates CTLs. The epitope is $N=$ 8--11 amino acids long, and in our
model we assume $N=9$. The initial strain of virus is denoted by $V_{0}$.
The assumption of an unique initial strain is based on the bottleneck
effect \cite{Manrubia} and the reversion effect due to lack of CTL pressure
at infection \cite{Kent}. We assume that at each of the $N$ sites on the
epitope there are $A$ possible amino acid substitutions. The quasispecies
space size is thus $I=( A+1)^{N}$. We use $A=2$ because each
site can accept few substitutions to keep the virus viable. We find that the
simulation results are not sensitive to $A$. The subscript of $V_{i}$ is in
the range $0\leq i\leq (1+A)^{N}-1$ and can be written in form of a $A+1$%
-radix number $i=(a_{N},a_{N-1},\cdots ,a_{1})$, with $0\leq a_{n}\leq A$.
The Hamming distance between epitopes is $d_{ij}=\sum_{n=1}^{N}(1-\delta
_{a_{n}^{i},a_{n}^{j}})$, where $\delta _{lm}$ is the Kronecker delta. The
model is
\begin{eqnarray}
\frac{dv_{i}}{dt}&=&r_{i}v_{i}-mNAv_{i}+m\Gamma _{i}-c_{1}f_{i}( \mathbf{x
}) v_{i}  \label{eq:dvi}
\\ 
\frac{dx_{i}}{dt}&=&\lambda \left( \frac{v_{i}}{v}-x_{i}\right)
-c_{3}g_{i}( \mathbf{x}) x_{i},  \label{eq:dxi}
\end{eqnarray}
for $i=0,1,\cdots I-1$, with the initial condition $v_{0}(0)\neq 0,v_{i\neq
0}(0)=0$, and \ $x_{i}(0)=0$. Previous models have assumed linear
suppression ($f_{i}=$ const) and not tracked both viral and T cell
quasispecies evolution \cite{Perelson}.

In Eq.\ (\ref{eq:dvi}), $r_{i}=r_{0}\exp ( -\alpha d_{i0}) $ is
the replication rate of $V_{i}$. The wild type $V_{0}$ replicates the
fastest, and mutation away demands some fitness cost. The term $m$ is the
rate of mutation of the amino acids at any one of the $N$ sites into any one
of the other $A$ amino acids (per amino acid per day). The term $mNAv_{i}$
thus represents mutation away from quasispecies $i$ to its $NA$\ neighbors
(efflux). The term $m\Gamma _{i}$ represents the reverse process (influx),
where
$
\Gamma _{i}=\sum_{j=0}^{A}
\left(
v_{(j,a_{N-1},\ldots, a_{1})}+v_{(a_{N},j,\ldots, a_{1})} 
+\ldots +v_{(a_{N},a_{N-1},\ldots, j)}
\right) -Nv_{i}
$
is the sum over all viral species a unit Hamming distance away from $i$. The
term
$
f_{i}( \mathbf{x}) =y_{i}( \mathbf{x}) / 
[ c_{2}+y_{i}( \mathbf{x}) ]
$
represents the lysis ratio of targeted cells harboring $V_{i}$, where
$
y_{i}( \mathbf{x})
=\sum_{j=0}^{I-1}x_{j}k_{ji}=\sum_{j=0}^{I-1}x_{j}\exp ( -\beta
d_{ji}) 
$
The derivation of 
$f_{i}( \mathbf{x})$
follows as in \cite{Park}
by treating the competitive binding process as a Langmuir adsorption
isotherm of the CTLs onto targeted cells that are infected with $V_{i}$. The
concentration and affinity here are dimensionless, while in \cite{Park}
their physical values $\{ x_{j}\} $ and $\{ k_{ji}\} $
are used.
That is,
$
f_{i}( \mathbf{x}) =\sum_{j}\{ x_{j}\} \{ k_{ji}\} 
/[
1+\sum_{j}\{ x_{j}\} \{ k_{ji}\}  ] 
$.
We let $\{ x \} _{\max }$ and $\{ k \} _{\max }$ denote the
maximum values of concentration and affinity, respectively.
We, thus, obtain $c_{2}=1/(
\{ x \} _{\max } \{ k \} _{\max }) $. Typical
concentrations are on the order of $10^{-6}$ mol/l, typical binding
constants are on the order of $10^{6}$ l/mol, and so we expect $c_{2}$ to be
on the order of unity. Indeed, to fit clinical data, we will find 
$c_{2}=0.8$. 
The coefficient $c_{1}$ is the rate of clearance of virus released from
targeted cells that are being lysed, and to fit clinical data, we will find $%
c_{1}=4.0$ day$^{-1}$, close to a previous estimation of 4.3 day$^{-1}$
\cite{Nelson}.

The first part of the right hand side of Eq.\ (\ref{eq:dxi}) represents a
delay process with time constant $1/\lambda $, since the immune system
always lags in response to viral changes. The term $v_{i}/v$, which equals $%
x_{i}$ in an ideal immune system, now serves as the moving target for $x_{i}$
to catch. Moreover, the tracking of the virus by the immune system is
disturbed by the competition $c_{3}g_{i}( \mathbf{x}) x_{i}$,
where
$
g_{i}( \mathbf{x} ) =\sum_{j\neq i}x_{j}k_{ji}=y_{i}( \mathbf{%
x} ) -x_{i} 
$
represents competition from the other T cells. This term has been used in
studying the competition within an ecosystem
\cite{Bastolla}. The T cells that develop into $X_{i}$%
, must be stimulated by $V_{i}$. However, since $X_{j}$ ($j\neq i$) can
occupy $V_{i}$, a delay in the maturation of the T cell precursors into $%
X_{i}$ can occur.

The number of non-linear equations in the model is $2(A+1)^{N}=39366$, which
is difficult for analysis and computation. However, there is apparent
symmetry in these equations: $v_{i}=v_{j}$ and $x_{i}=x_{j}$ as long as $%
d_{i0}=d_{j0}$. That is, all the viruses that have an identical distance to
the wild type have the same dynamics and can be put into the same group. We
let $\tilde{v}_{n}$ ($\tilde{x}_{n}$) denote the number of elements in the
viral (CTL) group $n$, one has $\tilde{v}_{n}=Q_{N}^{n}v_{i}$ and \ $\tilde{x%
}_{n}=Q_{N}^{n}x_{i}$ for any $i$ satisfying $d_{i0}=n$, where $%
Q_{N}^{n}=C_{N}^{n}A^{n}$ 
and $C_{N}^{n}=N!/[ n!(N-n)!] $.
 By
substituting $\tilde{v}_{n}$ and $\tilde{x}_{n}$ into Eqs.\ (\ref{eq:dvi}, %
\ref{eq:dxi}) and noting that
$
\Gamma _{i}=(n/Q_{N}^{n-1})\tilde{v}_{n-1}+
[n( A-1) /
Q_{N}^{n}]\tilde{v}_{n}+
[( N-n) A/Q_{N}^{n+1}]\tilde{v}_{n+1}
$,
one obtains
\begin{eqnarray}
\frac{d\tilde{v}_{n}}{dt} &=&mA\left( N-n+1\right) \tilde{v}_{n-1}+m\left(
n+1\right) \tilde{v}_{n+1}  \notag \\
&&+\left( \gamma _{0}\exp \left( -\alpha n\right) -c_{1}\tilde{f}_{n}\left(
\mathbf{\tilde{x}}\right) \right) \tilde{v}_{n}  \notag \\
&&-mA\left( N-n+n/A\right) \tilde{v}_{n}  \label{eq:dvn}
\\
\frac{d\tilde{x}_{n}}{dt}&=&\lambda \left( \frac{\tilde{v}_{n}}{v}-\tilde{x}%
_{n}\right) -c_{3}\tilde{y}_{n}\left( \mathbf{\tilde{x}}\right) \tilde{x}%
_{n}+\frac{c_{3}}{Q_{N}^{n}}\tilde{x}_{n}^{2}  \label{eq:dxn1}
\end{eqnarray}
for $n=0,1,\cdots ,N$, where $\tilde{f}_{n}( \mathbf{\tilde{x}}%
) =\tilde{y}_{n}( \mathbf{\tilde{x}}) /[c_{2}+\tilde{y}%
_{n}( \mathbf{\tilde{x}}) ]$. To obtain $\tilde{y}_{n}(
\mathbf{\tilde{x}}) $ from $y_{i}( \mathbf{x}) $, the
following two steps are performed.\nolinebreak\ \nolinebreak The first step
is to rearrange 
the equation for $y_i$ 
by summing up all the $x_{j}$ that have
the same $d_{ji}=n^{\prime }$, namely to obtain
$
\tilde{y}_{n}
( \mathbf{\tilde{x}}) =\sum_{n^{\prime
}=0}^{N}q_{n^{\prime }}\exp ( -\beta n^{\prime })$.
To be specific, let us consider
$
i=\left( \underset{N-n}{\underbrace{0\cdots 0}}\underset{n}{\underbrace{%
11\cdots 1}}\right)  
$
with $d_{0i}=n$ and a specific $d_{ji}=n^{\prime }$, $j$ must satisfy $%
n-n^{\prime }\leq d_{j0}\leq n+n^{\prime }$. To obtain a $j$\ with $%
d_{ji}=n^{\prime }$\ and $d_{j0}=k$, some 0's or 1's in this
expression for $i$
need to be changed, and the number of ways to realize this is $%
\sum_{l=k-n}^{n^{\prime }}Q_{N-n}^{l}C_{n}^{l_{1}}P_{n-l_{1}}^{l_{2}}$,
where $Q_{N-n}^{l}=C_{N-n}^{l}A^{l}$ corresponds to changing $l$ of the 0's
into non-0; $C_{n}^{l_{1}}$ corresponds to flipping $l_{1}$\ of the 1's into
0; $P_{n-l_{1}}^{l_{2}}=C_{n-l_{1}}^{l_{2}}( A-1) ^{l_{2}}$
corresponds to changing $l_{2}$\ of the remaining $(n-l_{1})$ 1's into
digits other than 0 and 1. The restrictions are $l+l_{1}+l_{2}=n^{\prime }$
and $n+l-l_{1}=k$, from which one solves $l_{1}=n+l-k$ and $l_{2}=n^{\prime
}-n+k-2l$. Therefore one obtains
$
q_{n^{\prime }}=\sum_{k=n-n^{\prime }}^{n+n^{\prime }}
(\tilde{x}_{k}/ Q_{N}^{k})\sum_{l=k-n}^{n^{\prime
}}Q_{N-n}^{l}C_{n}^{n+l-k}P_{k-l}^{n^{\prime }-n+k-2l}
$.
The second step is to sum up all terms that have
the same $\tilde{x}_{k}$. That is, to obtain the matrix $\Phi_{nk}$ 
that is defined by $\mathbf{\tilde{y}}=\Phi \mathbf{\tilde{x}}
$. By substituting the expression of $q_{n^{\prime }}$ into
the expression for $\tilde{y}_{n}$
one obtains
$
\Phi _{nk}=\sum_{n^{\prime }=0}^{N}
[\exp ( -\beta n^{\prime
}) /Q_{N}^{k}]\sum_{l=k-n}^{n^{\prime
}}Q_{N-n}^{l}C_{n}^{n+l-k}P_{k-l}^{n^{\prime }-n+k-2l}
$.

The model parameters are determined as follows. The term $mA$ is the
mutation rate per amino acid site per day. The conventional HIV-1 mutation
rate $\mu \approx 2.2\times 10^{-5}$ is per nucleotide base per replication
\cite{Huang}. Given that one replication cycle is about one day, one amino
acid has 3 nucleotide bases, and about $3/4$ nucleotide mutations are
nonsynonymous, one has $m=3\mu \times 0.75/A\approx 2.475\times 10^{-5}$ per
amino acid per day. The other parameter values are estimated by fitting the
standard data given by \cite{Pantaleo}. The viral load data is in terms of
the plasma HIV-1 titer, which is two raised to the number of serial $\times
1/2$ dilutions that result in undetectable virus. In the first viremia
phase, the mutation has not yet taken significant effect, Eqs.\ (\ref{eq:dvi},%
\ref{eq:dxi}) can be simplified as $\dot{v}_{0}=r_{0}v_{0}-c_{1}x_{0}v_{0}/%
( c_{2}+x_{0}) $ and $\dot{x}_{0}=\lambda ( 1-x_{0})$.
Since the CTL response peaks (i.e., $x_{0}$ approaching 1) at \ 9 to 12
weeks after the initial infection \cite{Abbas}, we calculate from $\dot{x_{0}%
}=\lambda (1-x_{0})$ that $\lambda $ is 0.033 day$^{-1}$. By noting that the
first viremia peaks roughly at week 6 with plasma virus titer reaching 630
and declines by week 12, we find $v_{0}( 0) =3.7 \times 10^{-7}$, $%
r_{0}=1.88$ day$^{-1}$, $c_{1}=4.0$, and $c_{2}=0.8$. The fitting of the
latency and the final viremia phases requires $\alpha =0.41$, $\beta =0.58$,
and $c_{3}=9.56\times 10^{-2}$. 
As discussed above, only 
$\alpha$ and $c_3$  lack comparison to literature values.

Eqs.\ (\ref{eq:dvn}, \ref{eq:dxn1}) are still difficult for analysis because $%
v$ shows up in the denominator of Eq.\ (\ref{eq:dxn1}). By summing 
Eq.\ (\ref{eq:dvn}) over $n$, one obtains
\begin{equation}
\frac{dv}{dt}=\sum_{n=0}^{N}\left( \gamma _{0}\exp ( -\alpha n)
-c_{1}\tilde{f}_{n}( \mathbf{\tilde{x}}) \right) \tilde{v}_{n}.
\label{eq:dv}
\end{equation}%
By defining $\tilde{p}_{n}=\tilde{v}_{n}/v$ and plugging $d\tilde{v}%
_{n}/dt=vd\tilde{p}_{n}/dt+\tilde{p}_{n}dv/dt$ into Eq.\ (\ref{eq:dvn}), one
obtains

\begin{eqnarray}
\frac{d\tilde{p}_{n}}{dt} &=&mA( N-n+1) \tilde{p}_{n-1}+  \notag
\\
&&\left( \gamma _{0}\exp ( -\alpha n) -c_{1}\tilde{f}_{n}
( \mathbf{\tilde{x}}) \right) \tilde{p}_{n}  \notag \\
&&-mA( N-n+n/A) \tilde{p}_{n}+m( n+1) \tilde{p}_{n+1}
\notag \\
&&-\tilde{p}_{n}\sum_{n^{\prime }=0}^{N}\tilde{p}_{n^{\prime }}
\left( \gamma _{0}\exp ( -\alpha n^{\prime }) -c_{1}\tilde{f}_{n^{\prime
}}( \mathbf{\tilde{x}}) \right)   \label{eq:dpdt}
\\
\frac{d\tilde{x}_{n}}{dt}&=&\lambda ( \tilde{p}_{n}-\tilde{x}_{n})
-c_{3}\tilde{y}_{n}( \mathbf{\tilde{x}}) \tilde{x}_{n}+\frac{c_{3}%
}{Q_{N}^{n}}\tilde{x}_{n}^{2},  \label{eq:dxdt}
\end{eqnarray}%
which are regular nonlinear ODEs. By integrating the above equations we find
that $\tilde{p}_{n}( t) $ and $\tilde{x}_{n}( t) $
approach the fixed point $(\bar{p}_{n}$, $\bar{x}_{n})$
as early as day 600. The eigenvalues of the Jacobian matrix evaluated
at the fixed point are all negative, indicating that the fixed point is
stable. That is, an equilibrium has been reached between the HIV-1
quasispecies and the immune system T cell quasispecies. This result holds
even as the viral load itself increases exponentially. All quasispecies
escape at the same rate, with their relative proportions fixed. To calculate
the rate of escape, Eq.\ (\ref{eq:dv}) is converted into $\dot{v}
( t) =\varepsilon ( t) v ( t) $, where
\begin{equation}
\varepsilon ( t) =\sum_{n=0}^{N}\tilde{p}_{n}\left( \gamma
_{0}\exp ( -\alpha n) -\frac{c_{1}\sum_{k}\Phi _{nk}\tilde{x}_{k}
}{c_{2}+\sum_{k}\Phi _{nk}\tilde{x}_{k}}\right) .  \label{eq:epsn}
\end{equation}
Because the viral quasispecies converges, $\varepsilon ( t) $\
will eventually converge to a constant $\bar{\varepsilon}$. Replacing $%
( \tilde{p}_{n},\tilde{x}_{n}) $ by the fixed value $( \bar{p%
}_{n},\bar{x}_{n}) $, one obtains $\bar{\varepsilon}=0.0066$ day$^{-1}$%
. Because the equilibrium begins early, we conclude that the majority of
latent phase, which was previously mysterious, follows the same simple
dynamics as the final viremia phase. This result explains the observation
\cite{Mellors}
that the lifespan of HIV patients is inversely correlated 
with the minimum viral load $%
v_{\min }$ that happens at about day 150 and is a direct result of the
primary immune response. Because the onset time to the final phase scales as
$t\propto -\ln v_{\min }/\bar{\varepsilon}$, the smallness of $\bar{%
\varepsilon}$ significantly amplifies the life span dependence on $v_{\min }$%
.

By incorporating immune competition,
we now have a physical theory based upon an analytical model. Upon
infection, the virus proliferates with a large rate around $1.88$ day$^{-1}$,
because the immune response has not been fully established. At about week
6, the immune system brings the virus down to a trace level.
However, the ever increasing viral genetic diversity induces competition
among CTLs, which degrades the quality of immune control. Both a viral and a
T cell quasispecies are established. The concentrations of different CTLs
become fixed, as well as the percentages of different viral quasispecies.
However, the virus escapes exponentially. Interestingly, it is this
underlying pair of quasispecies that drives the seemingly nonequilibrium
process. While small, $c_{3}$ has become the Achilles' heel of the immune
system. Competition among T cells may exert a small influence for many
diseases, but is fatal for HIV-1.

\begin{figure}[tbp]
\begin{center}
\epsfig{file=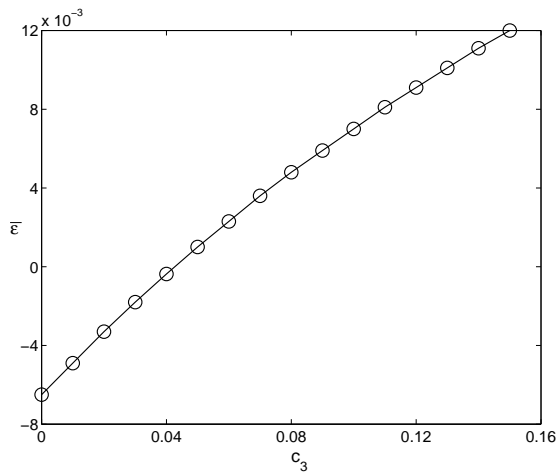,width=0.85\columnwidth,clip=}
\end{center}
\caption{The escape rate $\bar{\protect\varepsilon}$ increases as the
competition $c_3$ increases.}
\label{fig:epsilon}
\end{figure}

Our analysis predicts that competition among CTLs has a pivotal role on
viral dynamics. The function $\bar{\varepsilon}=\bar{\varepsilon}
( c_{3}) $ is shown in Fig. \ref{fig:epsilon}.
One sees that as $c_{3}$,
or the level of competition, increases, the escape rate $\bar{\varepsilon}$
increases significantly. For the ideal immune system, $c_{3}=0$ and $\bar{%
\varepsilon}=-0.0065$, suggesting that the ideal immune system can control
HIV-1 infection. 
Even for the real immune
system, the smallness of $\vert \bar{\varepsilon}\vert $ 
implies that HIV-1 and the immune system are well-matched competitors.
When all antigens are simultaneously present, competition 
among T cells for activation and division occurs, and $c_3$ is large.
By vaccinating to
different lymph nodes with the different HIV antigens likely to occur, 
competition between T cells is reduced, and T cells that recognize
each of the different antigens may be produced at high concentration
\cite{dengue}, and so $c_3 \approx 0$.
In this case the virus never rises again
after week 12. The regimen should be designed such that its components are
given according to the fixed viral quasispecies. According to our
simulation, the final viral percentages other than $\tilde{p}_{0}$ and $%
\tilde{p}_{5}$ are negligible. As for sensitivity to the parameters, if $%
\alpha $ or $\beta $ are increased, in which case the data of Fig.\ \ref%
{fig:timecourse} are not fit so well, $\tilde{p}_{6}$ also becomes
non-negligible. We thus propose a vaccine regimen that includes not only the
wild type but also the $\tilde{p}_{5}$ quasispecies. This vaccine must be
given before the eventual evolution of the quasispecies viral strains to
avoid immunodominance. If the vaccine were given after evolution of the $%
\tilde{p}_{5}$ species, then competition would be unavoidable due to prior
exposure of the immune system to the multiple viral strains.

\bibliography{hiv}

\end{document}